\documentclass[preprint,12pt]{elsarticle}
\usepackage{indentfirst}
\usepackage{bm}    
\usepackage{graphicx}
\usepackage[]{graphicx}  
\usepackage{enumerate}
\usepackage{cases}
\usepackage{amssymb}
\usepackage{mathrsfs}
\usepackage{bbm}
\usepackage{booktabs}
\usepackage{subfigure}
\makeatletter

\newcommand {\Rmnum} [1] {\expandafter \@slowromancap \romannumeral #1@}

\makeatletter

\newtheorem{Remark}{Remark}
\newtheorem{Protocol}{Protocol}

\begin{document}

\begin{frontmatter}

\title{A complete Classification of Quantum Public-key Encryption Protocols}

\author{Chenmiao Wu $^{1,2,3}$}
\author{Li Yang$^{1,2}$\corref{1}}

\cortext[1]{Corresponding author email: yangli@iie.ac.cn}
\address{1.State Key Laboratory of Information Security, Institute of Information Engineering, Chinese Academy of Sciences, Beijing 100093, China\\
2.Data Assurance and Communication Security Research Center,Chinese Academy of Sciences, Beijing 100093, China\\
3.University of Chinese Academy of Sciences, Beijing, 100049, China}


\begin{abstract}
We present a classification of quantum public-key encryption protocols. There are six elements in quantum public-key encryption: plaintext, ciphertext, public-key, private-key, encryption algorithm and decryption algorithm. According to the property of each element which is either quantum or classical, the quantum public-key encryption protocols can be divided into 64 kinds. Among 64 kinds of protocols, 8 kinds have already been constructed, 52 kinds can be proved to be impossible to construct and the remaining 4 kinds have not been presented effectively yet. This indicates that the research on quantum public-key encryption protocol should be focus on the existed kinds and the unproposed kinds.
\end{abstract}


\end{frontmatter}



\section{INTRODUCTION}
Diffie and Hellman \cite{1} lay the foundation of modern cryptography by presenting the concept of public-key encryption. Contrary to symmetric-key encryption which uses the same key to encrypt and to decrypt, public-key encryption uses a pair of different keys, public-key and private-key, to encrypt and to decrypt. Classical public-key encryption has been widely used in many fields. However, Shor's algorithm \cite{2} and Gover's algorithm \cite{3} can attack the computational difficulty problems which are used as the premises of constructing classical public-key encryption. The research on quantum public-key encryption can pre-empt such potential threats.

 Analogous to the analysis of quantum symmetric-encryption protocols\cite{19}, a quantum public-key encryption consists of six elements: plaintext, ciphertext, public-key, private-key, encryption algorithm and decryption algorithm. If an encryption protocol involved quantum part, it can be taken as the counterpart of classical public-key encryption. With the advent of quantum key distribution protocol, BB84\cite{4}, quantum public-key encryption has attained great development. Okamoto's protocol\cite{5} is considered as the first public-key encryption protocol which is based on the Knapsack-set problem. Gottesman\cite{6} firstly proposed "quantum public key cryptography with information-theoretic security". Up to now, there are some quantum public-key encryption protocols. For purpose of further analyzing the property of quantum public-key encryption and constructing new protocols, we classify the quantum public-key encryption protocols.

 This paper is organised as follows: a brief classification of quantum public-key encryption protocols is given in section 2; we classify these protocols into 3 types, each type is explained in section 3,4,5; in section 6, we discuss some problems exist in quantum public-key encryption protocols; finally, conclusions are drawn in section 7.
\section{Classification of quantum public-key encryption protocols}
A quantum public-key encryption protocol is a six-tuple $(P, C, PK, SK, \mathscr{E}, $ $\mathscr{D})$: plaintext $(P)$, ciphertext $(C)$, public-key $(PK)$, private-key $(SK)$, encryption algorithm $(\mathscr{E})$ and decryption algorithm $(\mathscr{D})$. Each element can be either classical or quantum, so there are 64 kinds of quantum public-key encryption protocols in total. Among them, 8 kinds have already been constructed, 52 kinds can be proved to be impossible to construct and the remaining 4 kinds have not been presented effectively yet. Thus, we can classify them into three types. The detail classification is show in table 1.
\label{sec:intro}  
\begin{table}[htbp]\footnotesize
  \hfil
  \caption{64 kinds of quantum public-key encryption protocols}
    \begin{tabular}{|c|cccccc|c|}
    \toprule
    {\bf Kind}      & ($P$     & $C$     & $PK$  & $SK$   & $\mathscr{E}$     & $\mathscr{D}$)   & existence  \\
    \midrule
    1     & $\mathbb{C}$ & $\mathbb{C}$ & $\mathbb{C}$ & $\mathbb{C}$ & $\mathbb{C}$ & $\mathbb{C}$ & $\mathbb{E}$ \\
    2     & $\mathbb{C}$ & $\mathbb{C}$ & $\mathbb{C}$ & $\mathbb{C}$ & $\mathbb{C}$ & $\mathbb{Q}$ & $\mathbb{E}$ \\
    3     & $\mathbb{C}$ & $\mathbb{C}$ & $\mathbb{C}$ & $\mathbb{C}$ & $\mathbb{Q}$ & $\mathbb{C}$ & $\mathbb{O}$ \\
    4     & $\mathbb{C}$ & $\mathbb{C}$ & $\mathbb{C}$ & $\mathbb{C}$ & $\mathbb{Q}$ & $\mathbb{Q}$ & $\mathbb{O}$ \\
    5     & $\mathbb{C}$ & $\mathbb{Q}$ & $\mathbb{C}$ & $\mathbb{C}$ & $\mathbb{C}$ & $\mathbb{C}$ & $\mathbb{N}$ \\
    6     & $\mathbb{C}$ & $\mathbb{Q}$ & $\mathbb{C}$ & $\mathbb{C}$ & $\mathbb{C}$ & $\mathbb{Q}$ & $\mathbb{N}$ \\
    7     & $\mathbb{C}$ & $\mathbb{Q}$ & $\mathbb{C}$ & $\mathbb{C}$ & $\mathbb{Q}$ & $\mathbb{C}$ & $\mathbb{N}$ \\
    8     & $\mathbb{C}$ & $\mathbb{Q}$ & $\mathbb{C}$ & $\mathbb{C}$ & $\mathbb{Q}$ & $\mathbb{Q}$ & $\mathbb{E}$ \\
    9     & $\mathbb{C}$ & $\mathbb{C}$ & $\mathbb{Q}$ & $\mathbb{C}$ & $\mathbb{C}$ & $\mathbb{C}$ & $\mathbb{N}$ \\
    10    & $\mathbb{C}$ & $\mathbb{C}$ & $\mathbb{Q}$ & $\mathbb{C}$ & $\mathbb{C}$ & $\mathbb{Q}$ & $\mathbb{N}$ \\
    11    & $\mathbb{C}$ & $\mathbb{C}$ & $\mathbb{Q}$ & $\mathbb{C}$ & $\mathbb{Q}$ & $\mathbb{C}$ & $\mathbb{N}$ \\
    12    & $\mathbb{C}$ & $\mathbb{C}$ & $\mathbb{Q}$ & $\mathbb{C}$ & $\mathbb{Q}$ & $\mathbb{Q}$ & $\mathbb{N}$ \\
    13    & $\mathbb{C}$ & $\mathbb{C}$ & $\mathbb{C}$ & $\mathbb{Q}$ & $\mathbb{C}$ & $\mathbb{C}$ & $\mathbb{N}$ \\
    14    & $\mathbb{C}$ & $\mathbb{C}$ & $\mathbb{C}$ & $\mathbb{Q}$ & $\mathbb{C}$ & $\mathbb{Q}$ & $\mathbb{N}$ \\
    15    & $\mathbb{C}$ & $\mathbb{C}$ & $\mathbb{C}$ & $\mathbb{Q}$ & $\mathbb{Q}$ & $\mathbb{C}$ & $\mathbb{N}$ \\
    16    & $\mathbb{C}$ & $\mathbb{C}$ & $\mathbb{C}$ & $\mathbb{Q}$ & $\mathbb{Q}$ & $\mathbb{Q}$ & $\mathbb{N}$ \\
    17    & $\mathbb{C}$ & $\mathbb{Q}$ & $\mathbb{Q}$ & $\mathbb{C}$ & $\mathbb{C}$ & $\mathbb{C}$ & $\mathbb{N}$ \\
    18    & $\mathbb{C}$ & $\mathbb{Q}$ & $\mathbb{Q}$ & $\mathbb{C}$ & $\mathbb{C}$ & $\mathbb{Q}$ & $\mathbb{N}$ \\
    19    & $\mathbb{C}$ & $\mathbb{Q}$ & $\mathbb{Q}$ & $\mathbb{C}$ & $\mathbb{Q}$ & $\mathbb{C}$ & $\mathbb{N}$ \\
    20    & $\mathbb{C}$ & $\mathbb{Q}$ & $\mathbb{Q}$ & $\mathbb{C}$ & $\mathbb{Q}$ & $\mathbb{Q}$ & $\mathbb{E}$ \\
    21    & $\mathbb{C}$ & $\mathbb{C}$ & $\mathbb{Q}$ & $\mathbb{Q}$ & $\mathbb{C}$ & $\mathbb{C}$ & $\mathbb{N}$ \\
    22    & $\mathbb{C}$ & $\mathbb{C}$ & $\mathbb{Q}$ & $\mathbb{Q}$ & $\mathbb{C}$ & $\mathbb{Q}$ & $\mathbb{N}$ \\
    23    & $\mathbb{C}$ & $\mathbb{C}$ & $\mathbb{Q}$ & $\mathbb{Q}$ & $\mathbb{Q}$ & $\mathbb{C}$ & $\mathbb{N}$ \\
    24    & $\mathbb{C}$ & $\mathbb{C}$ & $\mathbb{Q}$ & $\mathbb{Q}$ & $\mathbb{Q}$ & $\mathbb{Q}$ & $\mathbb{N}$ \\
    25    & $\mathbb{C}$ & $\mathbb{Q}$ & $\mathbb{C}$ & $\mathbb{Q}$ & $\mathbb{C}$ & $\mathbb{C}$ & $\mathbb{N}$ \\
    26    & $\mathbb{C}$ & $\mathbb{Q}$ & $\mathbb{C}$ & $\mathbb{Q}$ & $\mathbb{C}$ & $\mathbb{Q}$ & $\mathbb{N}$ \\
    27    & $\mathbb{C}$ & $\mathbb{Q}$ & $\mathbb{C}$ & $\mathbb{Q}$ & $\mathbb{Q}$ & $\mathbb{C}$ & $\mathbb{N}$ \\
    28    & $\mathbb{C}$ & $\mathbb{Q}$ & $\mathbb{C}$ & $\mathbb{Q}$ & $\mathbb{Q}$ & $\mathbb{Q}$ & $\mathbb{O}$ \\
    29    & $\mathbb{C}$ & $\mathbb{Q}$ & $\mathbb{Q}$ & $\mathbb{Q}$ & $\mathbb{C}$ & $\mathbb{C}$ & $\mathbb{N}$ \\
    30    & $\mathbb{C}$ & $\mathbb{Q}$ & $\mathbb{Q}$ & $\mathbb{Q}$ & $\mathbb{C}$ & $\mathbb{Q}$ & $\mathbb{N}$ \\
    31    & $\mathbb{C}$ & $\mathbb{Q}$ & $\mathbb{Q}$ & $\mathbb{Q}$ & $\mathbb{Q}$ & $\mathbb{C}$ & $\mathbb{N}$ \\
    32    & $\mathbb{C}$ & $\mathbb{Q}$ & $\mathbb{Q}$ & $\mathbb{Q}$ & $\mathbb{Q}$ & $\mathbb{Q}$ & $\mathbb{E}$ \\
    \bottomrule
    \end{tabular}%
     \begin{tabular}{|c|cccccc|c|}
    \toprule
    {\bf Kind}      & ($P$     & $C$     & $PK$  & $SK$   & $\mathscr{E}$     & $\mathscr{D}$)   & existence  \\
    \midrule
    33     & $\mathbb{Q}$ & $\mathbb{C}$ & $\mathbb{C}$ & $\mathbb{C}$ & $\mathbb{C}$ & $\mathbb{C}$ & $\mathbb{N}$ \\
    34     & $\mathbb{Q}$ & $\mathbb{C}$ & $\mathbb{C}$ & $\mathbb{C}$ & $\mathbb{C}$ & $\mathbb{Q}$ & $\mathbb{N}$ \\
    35     & $\mathbb{Q}$ & $\mathbb{C}$ & $\mathbb{C}$ & $\mathbb{C}$ & $\mathbb{Q}$ & $\mathbb{C}$ & $\mathbb{N}$ \\
    36    & $\mathbb{Q}$ & $\mathbb{C}$ & $\mathbb{C}$ & $\mathbb{C}$ & $\mathbb{Q}$ & $\mathbb{Q}$ & $\mathbb{N}$ \\
    37     & $\mathbb{Q}$ & $\mathbb{Q}$ & $\mathbb{C}$ & $\mathbb{C}$ & $\mathbb{C}$ & $\mathbb{C}$ & $\mathbb{N}$ \\
    38     & $\mathbb{Q}$ & $\mathbb{Q}$ & $\mathbb{C}$ & $\mathbb{C}$ & $\mathbb{C}$ & $\mathbb{Q}$ & $\mathbb{N}$ \\
    39     & $\mathbb{Q}$ & $\mathbb{Q}$ & $\mathbb{C}$ & $\mathbb{C}$ & $\mathbb{Q}$ & $\mathbb{C}$ & $\mathbb{N}$ \\
    40     & $\mathbb{Q}$ & $\mathbb{Q}$ & $\mathbb{C}$ & $\mathbb{C}$ & $\mathbb{Q}$ & $\mathbb{Q}$ & $\mathbb{E}$ \\
    41     & $\mathbb{Q}$ & $\mathbb{C}$ & $\mathbb{Q}$ & $\mathbb{C}$ & $\mathbb{C}$ & $\mathbb{C}$ & $\mathbb{N}$ \\
    42    & $\mathbb{Q}$ & $\mathbb{C}$ & $\mathbb{Q}$ & $\mathbb{C}$ & $\mathbb{C}$ & $\mathbb{Q}$ & $\mathbb{N}$ \\
    43    & $\mathbb{Q}$ & $\mathbb{C}$ & $\mathbb{Q}$ & $\mathbb{C}$ & $\mathbb{Q}$ & $\mathbb{C}$ & $\mathbb{N}$ \\
    44    & $\mathbb{Q}$ & $\mathbb{C}$ & $\mathbb{Q}$ & $\mathbb{C}$ & $\mathbb{Q}$ & $\mathbb{Q}$ & $\mathbb{N}$ \\
    45    & $\mathbb{Q}$ & $\mathbb{C}$ & $\mathbb{C}$ & $\mathbb{Q}$ & $\mathbb{C}$ & $\mathbb{C}$ & $\mathbb{N}$ \\
    46    & $\mathbb{Q}$ & $\mathbb{C}$ & $\mathbb{C}$ & $\mathbb{Q}$ & $\mathbb{C}$ & $\mathbb{Q}$ & $\mathbb{N}$ \\
    47    & $\mathbb{Q}$ & $\mathbb{C}$ & $\mathbb{C}$ & $\mathbb{Q}$ & $\mathbb{Q}$ & $\mathbb{C}$ & $\mathbb{N}$ \\
    48    & $\mathbb{Q}$ & $\mathbb{C}$ & $\mathbb{C}$ & $\mathbb{Q}$ & $\mathbb{Q}$ & $\mathbb{Q}$ & $\mathbb{N}$ \\
    49    & $\mathbb{Q}$ & $\mathbb{Q}$ & $\mathbb{Q}$ & $\mathbb{C}$ & $\mathbb{C}$ & $\mathbb{C}$ & $\mathbb{N}$ \\
    50    & $\mathbb{Q}$ & $\mathbb{Q}$ & $\mathbb{Q}$ & $\mathbb{C}$ & $\mathbb{C}$ & $\mathbb{Q}$ & $\mathbb{N}$ \\
    51    & $\mathbb{Q}$ & $\mathbb{Q}$ & $\mathbb{Q}$ & $\mathbb{C}$ & $\mathbb{Q}$ & $\mathbb{C}$ & $\mathbb{N}$ \\
    52    & $\mathbb{Q}$ & $\mathbb{Q}$ & $\mathbb{Q}$ & $\mathbb{C}$ & $\mathbb{Q}$ & $\mathbb{Q}$ & $\mathbb{E}$ \\
    53    & $\mathbb{Q}$ & $\mathbb{C}$ & $\mathbb{Q}$ & $\mathbb{Q}$ & $\mathbb{C}$ & $\mathbb{C}$ & $\mathbb{N}$ \\
    54    & $\mathbb{Q}$ & $\mathbb{C}$ & $\mathbb{Q}$ & $\mathbb{Q}$ & $\mathbb{C}$ & $\mathbb{Q}$ & $\mathbb{N}$ \\
    55    & $\mathbb{Q}$ & $\mathbb{C}$ & $\mathbb{Q}$ & $\mathbb{Q}$ & $\mathbb{Q}$ & $\mathbb{C}$ & $\mathbb{N}$ \\
    56    & $\mathbb{Q}$ & $\mathbb{C}$ & $\mathbb{Q}$ & $\mathbb{Q}$ & $\mathbb{Q}$ & $\mathbb{Q}$ & $\mathbb{N}$ \\
    57    & $\mathbb{Q}$ & $\mathbb{Q}$ & $\mathbb{C}$ & $\mathbb{Q}$ & $\mathbb{C}$ & $\mathbb{C}$ & $\mathbb{N}$ \\
    58    & $\mathbb{Q}$ & $\mathbb{Q}$ & $\mathbb{C}$ & $\mathbb{Q}$ & $\mathbb{C}$ & $\mathbb{Q}$ & $\mathbb{N}$ \\
    59    & $\mathbb{Q}$ & $\mathbb{Q}$ & $\mathbb{C}$ & $\mathbb{Q}$ & $\mathbb{Q}$ & $\mathbb{C}$ & $\mathbb{N}$ \\
    60    & $\mathbb{Q}$ & $\mathbb{Q}$ & $\mathbb{C}$ & $\mathbb{Q}$ & $\mathbb{Q}$ & $\mathbb{Q}$ & $\mathbb{O}$ \\
    61    & $\mathbb{Q}$ & $\mathbb{Q}$ & $\mathbb{Q}$ & $\mathbb{Q}$ & $\mathbb{C}$ & $\mathbb{C}$ & $\mathbb{N}$ \\
    62    & $\mathbb{Q}$ & $\mathbb{Q}$ & $\mathbb{Q}$ & $\mathbb{Q}$ & $\mathbb{C}$ & $\mathbb{Q}$ & $\mathbb{N}$ \\
    63    & $\mathbb{Q}$ & $\mathbb{Q}$ & $\mathbb{Q}$ & $\mathbb{Q}$ & $\mathbb{Q}$ & $\mathbb{C}$ & $\mathbb{N}$ \\
    64    & $\mathbb{Q}$ & $\mathbb{Q}$ & $\mathbb{Q}$ & $\mathbb{Q}$ & $\mathbb{Q}$ & $\mathbb{Q}$ & $\mathbb{E}$ \\
    \bottomrule
    \end{tabular}%
  \label{tab:classification}%

\end{table}%

In Table 1, $\mathbb{C}$ denotes the element belongs to classical space, $\mathbb{Q}$ denotes the element belongs to quantum space, $\mathbb{E}$ means the protocol exists, $\mathbb{N}$ means the protocol does not exist, $\mathbb{O}$ means whether this kind of protocol exists or not is still an open problem.
\section{Type E}
In this section, we will give simple examples about the existing 8 kinds of quantum public-key encryption protocol. The numbers of these protocols are 1, 2, 8, 20, 32, 40, 52, 64.
\subsection{Kind 1}
Broadly speaking, the classical public-key encryption protocol is a special case of quantum public-key encryption protocol. Each elements in Classical public-key encryption protocol is belong to classical space. There is no doubt that Kind 1 exists, such as RSA\cite{7} , ECC cryptosystem \cite{8} , Rabin cryptosystem \cite{9} and so on. Okamoto proposed a QPKE protocol \cite{5} based on the Knapsack-set problem. All the six elements of this protocol are classical, but the participants are all quantum Turing machines.
\subsection{Kind 2}
This kind uses quantum computation to decrypt. Supposed that during the decryption of ElGamal cryptosystem \cite{10} , the participant can use Shor's algorithm to solve the discrete problem to get the transmitted message. Quantum attack to the classical PKE protocol is also belong to this kind.
\subsection{Kind 8}
$P, PK, SK\in\mathbb{C}$ and $C, \mathscr{E}, \mathscr{D}\in\mathbb{Q}$ according to the kind 8. We can use the QPKE\cite{11} based on the McEliece cryptosystem \cite{12} to transmit classical message $x$.
\begin{Protocol}[$P, PK, SK\in\mathbb{C};C, \mathscr{E}, \mathscr{D}\in\mathbb{Q}$]

\rm{Let Alice's public-key be $G^{'}$, $G^{'}=SGP$. $S$ is an invertible matrix, $P$ is a permutation matrix and $G$ is a generator matrix of a Goppa code. And let her private-key be $(S, G, P)$.}
 \rm{ \begin{bfseries}
\flushleft{[Encryption]}
\end{bfseries}}
Bob uses Alice's public-key $G^{'}$ to encrypt the classicla message $x$ as below:
\begin{enumerate}
\item he prepares a quantum sate $|x\rangle$ and does Hadamard transformation on it:
\begin{eqnarray}
|x\rangle&\rightarrow&\frac{1}{\sqrt{2}}\sum\limits_{m}(-1)^{m\cdot x}|m\rangle\nonumber\\
&=&\sum\limits_{m}\alpha_{m}|m\rangle
\end{eqnarray}
\item he does transformation on the quantum message with Alice's public-key:
\begin{eqnarray}
U_{G^{'}}(\sum\limits_{m}\alpha_{m}|m\rangle_{k}|0\rangle_{n})=\sum\limits_{m}\alpha_{m}|m\rangle_{k}|mG^{'}\rangle_{n};
\end{eqnarray}

\item he randomly chooses an error $e$ to add into the state:
\begin{eqnarray}
U_{e}(\sum\limits_{m}\alpha_{m}|mG^{'}\rangle_{n})=\sum\limits_{m}\alpha_{m}|mG^{'}\oplus e\rangle_{n},
\end{eqnarray}
and then sends $\sum\limits_{m}\alpha_{m}|mG^{'}\oplus e\rangle_{n}$ to Alice.
\end{enumerate}
 \rm{ \begin{bfseries}
\flushleft{[Decryption]}
\end{bfseries}}
Alice uses her private-key $(S, G, P)$ to decrypt the receiving ciphertext:
\begin{enumerate}
\item she uses $P^{-1}$ to do computation:
\begin{eqnarray}
U_{P^{-1}}(\sum\limits_{m}\alpha_{m}|m\rangle_{k}|mG^{'}\oplus e\rangle_{n})=\sum\limits_{m}\alpha_{m}|m\rangle_{k}|mSG\oplus e\rangle_{n};
\end{eqnarray}
\item she takes advantage of $H$ to apply the operator $U_{H}$:
\begin{eqnarray}
U_{H}(\sum\limits_{m}\alpha_{m}|mSG\oplus e\rangle_{n}|0\rangle_{n-k})=\sum\limits_{m}\alpha_{m}|mSG\oplus e\rangle_{n}|s\rangle_{n-k},
\end{eqnarray}
and then measures the second register to get the syndrome to find the error via the fast decoding algorithm of the Goppa code generated by G;
\item she uses error vector $e$ to do computation:
\begin{eqnarray}
U_{e}(\sum\limits_{m}\alpha_{m}|mSG\oplus e\rangle_{n})=\sum\limits_{m}\alpha_{m}|mSG\rangle_{n};
\end{eqnarray}
and also uses the inver matrix of $G$ to do transform:
\begin{eqnarray}
U_{G^{-1}}(\sum\limits_{m}\alpha_{m}|mSG\rangle_{n}|0\rangle_{k})=\sum\limits_{m}\alpha_{m}|mSG\rangle_{n}|mSGG^{-1}\rangle_{k};
\end{eqnarray}
\item finally, she does computation according to $S^{-1}$ to get the quantum message:
\begin{eqnarray}
U_{S^{-1}}(\sum\limits_{m}\alpha_{m}|mS\rangle_{k})&=&\sum\limits_{m}\alpha_{m}|mSS^{-1}\rangle_{k}\nonumber\\
&=&\sum\limits_{m}\alpha_{m}|m\rangle_{k};
\end{eqnarray}
\end{enumerate}

\end{Protocol}
\subsection{Kind 20}
Kind 20 represents that $P, SK\in\mathbb{C}$ and $C, PK, \mathscr{E}, \mathscr{D}\in\mathbb{Q}$. This kind consists of two subclasses, one is that the public-key is composed only by quantum state, the other is that it employs a random changed pair of classical string and quantum state as public-key.
\begin{Protocol}[$P, SK\in\mathbb{C}; C, PK, \mathscr{E}, \mathscr{D}\in\mathbb{Q}$]~~~~~~~~~~~~~~~~~~~~~~~~~~~~~~~~~~~~~~~~~~

\rm{In Kawachi's protocol\cite{17}, the public-key is the quantum state $\rho^{+}_{\pi}=|\sigma\rangle+|\sigma\pi\rangle$, where $\sigma\in S_{n}$. And the private-key is a permutation $\pi$ selected from $S_{n}$.}
\rm{ \begin{bfseries}
\flushleft{[Encryption]}
\end{bfseries}}
\begin{enumerate}
\item If Bob intends to send message "0" to Alice, he sends Alice $\rho^{+}_{\pi}$,
\item If he wants to transmit "1", he sends Alice $\rho^{-}_{\pi}$, where $\rho^{-}_{\pi}$ is obtained from the operations as follows:
\begin{eqnarray}
|\sigma\rangle+|\sigma\pi\rangle\mapsto(-1)^{sgn(\sigma)}|\sigma\rangle+(-1)^{sgn(\sigma\pi)}|\sigma\pi\rangle
\end{eqnarray}
\end{enumerate}
\rm{ \begin{bfseries}
\flushleft{[Decryption]}
\end{bfseries}}
Alice decrypts Bob's ciphertext using the private-key $\pi$.
\end{Protocol}
\begin{Protocol}[$P, SK\in\mathbb{C}; C, PK, \mathscr{E}, \mathscr{D}\in\mathbb{Q}$]~~~~~~~~~~~~~~~~~~~~~~~~~~~~~~~~~~~~~~~~~~

\rm{Quantum public-key encryption protocol\cite{18} based on conjugate coding employs a random changed pair of bit string and quantum state as public-key and a Boolean function as private-key. Let the public-key be $(s, H^{k}|i\rangle)$, the private-key be $F$, where $s$ is used just once time, $i\in \Omega_{b}$, $\Omega_{b}=\{i\in\{0,1\}^{n}|i_{1}\oplus\cdots\oplus i_{n}=b\}$. The owner of private-key uses bit string of public-key $s$ as input to $F$ to compute $k$ which is an important element in preparing quantum state.}
\rm{ \begin{bfseries}
\flushleft{[Encryption]}
\end{bfseries}}
\begin{enumerate}
\item Bob wants to transmit one bit message b to Alice, he gets one of Alice's public-key and then selects a bit $j$ from $\Omega_{b}$,
\item He applies $Y^{j}$ to $H^{k}|i\rangle$,
\item He sends $(s, Y^{j}H^{k}|i\rangle)$ to Alice.
\end{enumerate}
 \rm{ \begin{bfseries}
\flushleft{[Decryption]}
\end{bfseries}}
\begin{enumerate}
\item Alice calculates $F(s)=k$,
\item She removes the transformation $H^{k}$ from the quantum state, and gets $Y^{j}|i\rangle$,
\item She measures $Y^{j}|i\rangle$ with basis $\{0,1\}^{n}$ to acquire the plaintext.
\end{enumerate}
\end{Protocol}
\subsection{Kind 32}
This kind requests that $P\in\mathbb{C}$ and $C, PK, SK, \mathscr{E}, \mathscr{D}\in\mathbb{Q}$. Kind 32 usually uses EPR pairs as public-key and private-key\cite{13,14} .
\begin{Protocol}[$P\in\mathbb{C}; C, pK, sK, \mathscr{E}, \mathscr{D}\in\mathbb{Q}$]~~~~~~~~~~~~~~~~~~~~~~~~~~~~~~~~~~~~~~~~~~

\rm{Denote the public-key as $S_{p}=\{p_{1}, p_{2}, \ldots, p_{n}\}$ and private-key as $S_{q}=\{q_{1}, q_{2}, \ldots, q_{n}\}$. Both the public-key and private-key are in the Bell state: $|\Phi^{+}\rangle_{12}=\frac{1}{\sqrt{2}}(|00\rangle_{12}+|11\rangle_{12})$}. Particle 1 is belong to public-key and particle 2 is private-key.
 \rm{ \begin{bfseries}
\flushleft{[Encryption]}
\end{bfseries}}
If Bob intends to send r-bit classical message $m=\{m_{1}, m_{2}, \ldots, m_{r}\}$:
\begin{enumerate}
\item he prepares r-qubit quantum state $|m\rangle=\{|m_{1}\rangle, |m_{2}\rangle, \ldots, |m_{r}\rangle\}$,
\item he uses Alice's public-key $S_{p}$ to do CNOT operations on the prepared quantum state:
\begin{eqnarray}
C_{p_{i}l_{i}}|\Phi^{+}\rangle_{p_{i}q_{i}}|m_{i}\rangle_{l_{i}}=\frac{1}{\sqrt{2}}(|00m_{i}\rangle+|11\overline{m_{i}}\rangle)_{p_{i}q_{i}l_{i}},
\end{eqnarray}
\item sends all these particles to Alice.
\end{enumerate}
 \rm{ \begin{bfseries}
\flushleft{[Decryption]}
\end{bfseries}}
\begin{enumerate}
\item When receiving ciphertext, Alice uses her private-key $S_{q}$ to do CNOT operations:
\begin{eqnarray}
C_{q_{i}l_{i}}\frac{1}{\sqrt{2}}(|00m_{i}\rangle+|11\overline{m_{i}}\rangle)_{p_{i}q_{i}l_{i}}=|\Phi^{+}\rangle_{p_{i}q_{i}}|m_{i}\rangle_{l_{i}},
\end{eqnarray}
\item Alice measures the state with basis $\{|0\rangle, |1\rangle\}$ to get the plaintext.
\end{enumerate}
\end{Protocol}

\subsection{Kind 40}
This kind aims at encrypting quantum message, requiring $PK, SK\in\mathbb{C}$ and $P, C, \mathscr{E}, \mathscr{D}\in\mathbb{Q}$. Li Yang and Fujita proposed some quantum counterparts of McEliece cryptosystem\cite{11, 15} . Such protocols are belong to this kind. We can use Protocol 1 to transmit quantum message, so this kind exists.

\subsection{Kind 52}
This kind requersts: $SK\in\mathbb{C}$ and $P, C, PK, \mathscr{E}, \mathscr{D}\in\mathbb{Q}$.
Gotteseman was the first one to consider quantum state as public-key, and proposed an information-theoretically secure QPKE\cite{6}.
\begin{Protocol}[$sK\in\mathbb{C};P, C, pK, \mathscr{E}, \mathscr{D}\in\mathbb{Q}$]~~~~~~~~~~~~~~~~~~~~~~~~~~~~~~~~~~~~~~~~~~~~~~~~~~~~~~~~~~~~

\rm{Let Alice's public-key be $|\psi\rangle=\frac{1}{\sqrt{2}}(I\otimes U_{k})(|0\rangle|0\rangle+|1\rangle|1\rangle)$, and private-key be $k$.}
\rm{ \begin{bfseries}
\flushleft{[Encryption]}
\end{bfseries}}~~~~~~~~~~~~~~~~~~~~~~~~~~~~~~~~~~~~~~~~~~~~~~~~~~~~~~~~~~~~~~~~~~~~~~~~~~~~~~~~~~~~~~~~~

Bob uses Alice's public-key to encrypt quantum menssage $|\varphi\rangle=\alpha|0\rangle+\beta|1\rangle$:

he teleports the quantum message through the public-key:
\begin{eqnarray}
|\varphi\rangle|\psi\rangle&=&\frac{1}{\sqrt{2}}(\alpha|0\rangle+\beta|1\rangle)(|0\rangle\otimes U_{k}|0\rangle+|1\rangle\otimes U_{k}|1\rangle)
\end{eqnarray}
and by measuring the first and second particles, the whole system becomes:
\begin{eqnarray}
|\varphi\rangle|\psi\rangle&=&\frac{1}{2}|\Phi^{+}\rangle\big(\alpha U_{k}|0\rangle+\beta U_{k}|1\rangle\big)+\frac{1}{2}|\Phi^{-}\rangle\big(\alpha U_{k}|0\rangle-\beta U_{k}|1\rangle\big)+\nonumber\\
&&\frac{1}{2}|\Psi^{+}\rangle\big(\beta U_{k}|0\rangle+\alpha U_{k}|1\rangle\big)\frac{1}{2}|\Psi^{-}\rangle\big(-\beta U_{k}|0\rangle+\alpha U_{k}|1\rangle\big)\nonumber\\
&=&\frac{1}{2}|\Phi^{+}\rangle U_{k}\circ I|\varphi\rangle+\frac{1}{2}|\Phi^{-}\rangle U_{k}\circ Z|\varphi\rangle\nonumber\\
&&+\frac{1}{2}|\Psi^{+}\rangle U_{k}\circ X|\varphi\rangle+\frac{1}{2}|\Psi^{-}\rangle U_{k}\circ XZ|\varphi\rangle.
\end{eqnarray}

Bob gets the corresonding Pauli matrix $P$ and sends Alice $P$ and 2nd register of public-key.

\rm{ \begin{bfseries}
\flushleft{[Decryption]}
\end{bfseries}}~~~~~~~~~~~~~~~~~~~~~~~~~~~~~~~~~~~~~~~~~~~~~~~~~

After receiving ciphertext $(P, U_{k}P|\varphi\rangle)$, Alice decrypts it by performing $U^{-1}_{k}$ then $P^{-1}$, and gets the plaintext $|\varphi\rangle$.
\end{Protocol}
\begin{Remark}
\rm{The kind 52 can be extended to a multi-qubit protocol with the help of the multi-qubit unknown state teleportation, which can be accomplished in a qubit-wise way. The feasibility of this kind of teleportation can be proved with mathematical induction\cite{WuYang}.}

~~~~~~~~~~~~~~~~~~~~~~~~~~~~~~~~~~~~~

\rm{The teleportation of two-qubit state has been studied in the Ref. \cite{rigolin2005quantum,20,21}. Now, we suppose $(n-1)$-qubit state: $|\psi_{1}\rangle=\sum\limits_{i_{1},\ldots,i_{n-1}}\alpha_{i_{1},\ldots,i_{n-1}}|i_{1},\ldots,i_{n-1}\rangle$ has been already transmitted successfully by the same way of sending two-qubit state. And the operations to reconstruct the $(n-1)$ particles are $U_{1}\otimes\cdots\otimes U_{n-1}$. Now, we prove that the quantum state of $n$-qubit state will be send by n EPR pairs $|\Phi^{+}\rangle$.

The arbitrary n-qubit unknown state is a mixed state, which is the superposition of each component: $|\psi_{2}\rangle=\sum\limits_{i_{1}\ldots i_{n}}\alpha_{i_{1}\ldots i_{n}}|i_{1}\ldots i_{n}\rangle_{1,\ldots,n}$. The teleportation of $|\psi_{2}\rangle$ is equal to teleport each component. The mixed state is also pure state, and each component is independent. Thus, $|\psi_{2}\rangle$ can be expressed as: $|\psi_{2}\rangle=|\psi_{21}\rangle|0\rangle+|\psi_{22}\rangle|1\rangle$, where $|\psi_{21}\rangle$ and $|\psi_{22}\rangle$ are independent of each other. So another expression of $|\psi_{2}\rangle$ is:
\begin{eqnarray}
|\psi_{2}\rangle&=&|\psi_{21}\rangle_{1,\ldots,n-1}|0\rangle_{n}+|\psi_{22}\rangle_{1,\ldots,n-1}|1\rangle_{n}\nonumber\\
&=&\sum\limits_{i^{'}_{1}\ldots i^{'}_{n-1}}\alpha^{'}_{i^{'}_{1}\ldots i^{'}_{n-1}}|i^{'}_{1}\ldots i^{'}_{n-1}\rangle_{1,\ldots,n-1}|0\rangle_{n}+\nonumber\\
&&\sum\limits_{i^{''}_{1}\ldots i^{''}_{n-1}}\alpha^{''}_{i^{''}_{1}\ldots i^{''}_{n-1}}|i^{''}_{1}\ldots i^{''}_{n-1}\rangle_{1,\ldots,n-1}|1\rangle_{n}.
\end{eqnarray}
Alice and Bob preshare EPR pair $|\Phi^{+}\rangle_{n+1,n+2},\ldots,|\Phi^{+}\rangle_{3n-1,3n}$. Based on the assumption that $(n-1)$-qubit state is transmitted by the same method of qubit-wise teleportation, Bob also sends the nth qubit through teleportation. Since the mixed state can be considered as pure state, we conclude the previously transmitted $(n-1)$ qubits in the the following operation to simplify expression:
$ \sum\limits_{i_{1}\ldots i_{n}}\alpha_{i_{1}\ldots i_{n}}|i_{1}\ldots i_{n}\rangle_{1,\ldots,n}|\Phi^{+}\rangle_{3n-1,3n}$

    If Bob performs Bell measurement on particle $n$ and particle $3n-1$, his measurement outcome should be one of the four possibilities: $|\Phi^{+}\rangle_{n,3n-1}$, $|\Phi^{-}\rangle_{n,3n-1}$, $|\Psi^{+}\rangle_{n,3n-1}$ and $|\Psi^{-}\rangle_{n,3n-1}$. The probability of each result is $\frac{1}{4}$. Once Alice receives the message send by Bob, she can fix up her state, recovering $|\psi_{2}\rangle$, by applying the appropriate quantum gate $U_{1}\otimes\cdots\otimes U_{n}$. To each operation $U_{i}$, if Bob's measurement outcome yields $|\Phi^{+}\rangle$, Alice does not need to do anything. If the measurement outcome is $|\Phi^{-}\rangle$ then Alice can fix up her state by applying the $Z$ gate. If the measurement outcome is $|\Psi^{+}\rangle$, Alice does transformation $X$ to reconstruct the state. If $|\Psi^{-}\rangle$, Alice uses first an $X$ and then a $Z$ gate to recover the state.}
\end{Remark}
\subsection{Kind 64}
All the six elements in this kind are belong to quantum space, we give a simple example as follows:
\begin{Protocol}[$P, C, PK, SK, \mathscr{E}, \mathscr{D}\in\mathbb{Q}$]~~~~~~~~~~~~~~~~~~~~~~~~~~~~~~~~~~~

\rm{Let plaintext be $\sum\limits_{m}\alpha_{m}|m\rangle$. Public-key $S_{PK}$ and private-key $S_{SK}$ are in the entangled state
$\frac{1}{\sqrt{2}}(|0\rangle_{A}|0\rangle_{B}+|1\rangle_{A}|1\rangle_{B})$.}
\rm{ \begin{bfseries}
\flushleft{[Encryption]}
\end{bfseries}}
\begin{enumerate}
\item Bob performs CNOT operation on the quantum state with Alice's public-key $S_{pk}$:
\begin{eqnarray}
C_{p_{i}l_{i}}\sum\limits_{m}\alpha_{m}|\Phi^{+}\rangle_{p_{i}q_{i}}|m_{i}\rangle_{l_{i}}=\frac{1}{\sqrt{2}}(|00m_{i}\rangle+|11\overline{m_{i}}\rangle)_{p_{i}q_{i}l_{i}},
\end{eqnarray}
\item Bob sends the ciphertext to Alice.
\end{enumerate}
\rm{ \begin{bfseries}
\flushleft{[Decryption]}
\end{bfseries}}
\begin{enumerate}
\item Alice decrypts with her private-key $S_{sK}$:
\begin{eqnarray}
C_{q_{i}l_{i}}\frac{1}{\sqrt{2}}(|00m_{i}\rangle+|11\overline{m_{i}}\rangle)_{p_{i}q_{i}l_{i}}=|\Phi^{+}\rangle_{p_{i}l_{i}}|m_{i}\rangle_{l_{i}},
\end{eqnarray}
\item Alice gets the plaintext in the second register.
\end{enumerate}
\end{Protocol}
\section{Type N}
As shown in Table 1,  52 kinds protocol are unable to be constrcuted. The encryption expression of QPKE can be written as:
\begin{eqnarray}
C=Enc_{PK}(P).
 \end{eqnarray}
 It refers to that giving plaintext and public-key to the encryption algorithm, the output is the ciphertext.

 The expression of decryption is:
 \begin{eqnarray}
 P=Dec_{SK}(C).
 \end{eqnarray}
 The input for the decryption algorithm are ciphertext and private-key, and the plaintext will be obtained.

 According to Eq. 18, if one element of both the input and output is belong to quantum space, the encryption algorithm must have the ability to process quantum elements. So the encryption algorithm should be belong to quantum space. Through this analysis, we can confirm that kind 5, 6, 9, 10, 17, 18, 21, 22, 25, 26, 29, 30, 33, 34, 37, 38, 41, 42, 45, 46, 49, 50, 53, 54, 57, 58, 61, 62 does not exist.

 Analogous to encryption process, if one of private-key, ciphertext and plaintext is belong to quantum space, the decryption algorithm is also belong to quantum space. So kind 7, 11, 15, 19, 23, 27, 31, 35, 39, 43, 47, 51, 55, 59, 63 are impossible to construct.

 When one of the inputs for the quantum encryption algorithm, plaintext or public-key, the output should be quantum states. If not, it means Quantum coherent components disappear after encryption and the output remains only one component. So quantum state degenerates into a classical one. That is, most of the information in plaintext are missing with the processing of encryption. Such protocol is useless in transmitting message. Kind 12, 24, 36, 44, 48, 56 are non-exist.

 Similarly, if either ciphertext or private-key is quantum during the process of decryption, the plaintext must be quantum states. The reason is the same with encryption processing. Thus, kind 13, 14, 16 are impossible.

\section{Type O}
For Kind 3: $(P, C, PK, SK, \mathscr{D}\in\mathbb{C}; \mathscr{E}\in\mathbb{Q})$, Kind 4: $(P, C, PK, SK\in\mathbb{C}; \mathscr{E}, \mathscr{D}\in\mathbb{Q})$, Kind 28: $(P, PK,\in\mathbb{C}; C, SK, \mathscr{E}, \mathscr{D}\in\mathbb{Q})$ and Kind 60: $(PK,\in\mathbb{C};P, C, SK, \mathscr{E}, \mathscr{D}\in\mathbb{Q})$, we are uncertain about their existence and also cannot determin their non-existence. If we consider classical plaintext as a special case of quantum plaintext, Kind 28 can be classified into Kind 60. There may be some novel and intersting QPKE protocols of Kind 3, 4, 28, 60 which are worth further studying.
\section{Discussion}
By classification, we review some QPKE protocols that are fully studied, make a list of impossible QPKE protocols and figure out the kinds which their existence is still an open problem. However, when it comes to type O, we cannot either give a proof of its existence or of its non-existence. In this paper, public-key distribution and the key generation are not taken into consideration. If these two phase are considered in the classification of QPKE protocols, the problem would be more complex, and there may be more interesting QPKE protocols to think about.

In public-key encryption protocol, there are some security notions, such as computational security, information-theoretically security, semantic security, malleable, non-malleable and so on. If we classify the QPKE protocols according the security notions, the classification will be more complex but more specific.
\section{Conclusions}
Based on the six-tuple of QPKE, we classify the 64 kinds of QPKE protocols into three types. First type includes 8 kinds of protocols that all have been presented yet and worth deeply studying. Second type includes 52 kinds of protocols that are are proved to be unable to construct or are no use to construct. The remaining protocols are worth further discussing.
\section*{Acknowledgement}   

This work was supported by the National Natural Science Foundation of China under Grant No. 61173157.


\begin{thebibliography}{99}
\bibitem{1}W. Diffie and M. Hellman, New directions in cryptography, {\em IEEE Transactions on Information Theory}, {\bf 22}, 644-654, 1976.
\bibitem{2}P. W. Shor, Polynomial-time algorithms for prime factorization and discrete logarithms on a quantum computer, \emph{SIAM journal on computing},\textbf{26}(5): 1484-1509, 1997.
\bibitem{3}L. K. Grover, A fast quantum mechanical algorithm for database search, \emph{Proceedings of the twenty-eighth annual ACM symposium on Theory of computing}, ACM, 212-219, 1996.
    \bibitem{19}C. Xiang and L. Yang, The classification of quantum symmetric-encryption protocols, SPIE Asia Photonics 2014 , 2014.
\bibitem{4}C. H. Bennett and Brassard G. An update on quantum cryptography,\emph{ Advances in cryptology}. Springer Berlin Heidelberg, 475-480, 1985.
\bibitem{5}T. Okamoto, K. Tanaka and S. Uchiyama, Quantum public-key cryptosystems, Advances in Cryptology-Crypto 2000, LNCS 1880, 147-165, 2000.
\bibitem{6}D. Gottesman, Quantum public key cryptography with information-theoretic security, unpulished, 2005.
\bibitem{7}R. L. Rivest, A. Shamir and L. Adleman. A method for obtaining digital signatures and public-key cryptosystems, \emph{Communications of the ACM}, \textbf{21}(2): 120-126, 1978.
\bibitem{8}N. Koblitz, Elliptic curve cryptosystems, Mathematics of computation, \textbf{48}(177): 203-209, 1987.
\bibitem{9}M. O. Rabin, Digitalized signatures and public-key functions as intractable as factorization, MASSACHUSETTS INST OF TECH CAMBRIDGE LAB FOR COMPUTER SCIENCE, 1979.
\bibitem{10}R. Merkle and Hellman M E. Hiding information and signatures in trapdoor knapsacks, IEEE Transactions on Information Theory, \textbf{24}(5): 525-530, 1978.
\bibitem{11}L. Yang, Quantum Public-key Cryptosystem Based on Classical NP-Complete Problem, arXiv: quant-ph/0310076.
\bibitem{12}R. J. McEliece, A public-key cryptosystem based on algebraic coding theory, DSN progress report, \textbf{42}(44): 114-116, 1978.
\bibitem{17}A. Kawachi, T. Koshiba and H. nishimura, et al. Computational indistinguishability between quantum states and its cryptographic application, Advances in Cryptology-EUROCRYPT 2005: 268-284, Springer Berlin Heidalberg, 2005.
\bibitem{18}L. Yang, B. Y. Yang and C. Xiang, Quantum Public-Key Encryption Schemes Based on Conjugate Coding, arXiv: 1112.0421.
\bibitem{13}F. Gao, Q. Y. Wen, S. J. Qin, et al, Quantum asymmetric cryptography with symmetric keys, \emph{Sci China Ser G}, \textbf{52}(12): 1925-1931, 2009.
\bibitem{14}X. Li and D. Zhang, Quantum Public-Key Cryptosystem Based on Super Dense Coding Technology, \emph{Journal of Computers}, \textbf{8}(12): 3168-3175, 2013.
\bibitem{15}H. Fujita, Quantum McEliece public-key cryptosystem, \emph{Quantum Information and Computation}, \textbf{12}(3-4): 181-202, 2012.

\bibitem{WuYang}C. M. Wu and L. Yang, Qubit-wise teleportation and its application in public-key secret communication, submitted to arXivin in Apr 30, 2015.
\bibitem{rigolin2005quantum}
G. Rigolin, Quantum teleportation of an arbitrary two-qubit state and its relation to multipartite entanglement[J]. Physical Review A, 71(3): 032303, 2005.
\bibitem{20} H. P. Zhu, Perfect teleportation of an Arbitrary two-qubit state via GHZ-like states, \emph{Int J Theor Phys}, \textbf{53}: 4095-4097, 2014.
    \bibitem{21}C. H. Bennett, Purification of noisy entanglement and faithful teleportation via noisy channels, \emph{Phys. Rev. Lett,}
76: 722-725, 1996.
\bibitem{16}M. Liang and L. Yang, Public-key encryption and authentication of quantum information, \emph{Sci China-Phys Mech Astron}, \textbf{55}(9): 1618¨C1629, 2012,
\end{thebibliography}

\end{document}